\documentclass[prl,twocolumn,showpacs,superscriptaddress]{revtex4}

\usepackage[usenames]{color}
\usepackage{epsf}
\usepackage{epsfig}
\usepackage{graphicx}
\usepackage{latexsym}
\usepackage{bm}
\begin{document}

\title{Fragmentation dynamics of CO$_{2}^{3+}$ investigated by multiple electron capture in collisions with slow highly charged ions}

\author{\firstname{N.} \surname{Neumann}}
\affiliation{Institut f\"ur Kernphysik, Goethe-Universit\"at Frankfurt am Main, Max-von-Laue-Str.1, 60438 Frankfurt, Germany}

\author{\firstname{D.} \surname{Hant}}
\affiliation{Institut f\"ur Kernphysik, Goethe-Universit\"at Frankfurt am Main, Max-von-Laue-Str.1, 60438 Frankfurt, Germany}

\author{\firstname{L.Ph.H.} \surname{Schmidt}}
\affiliation{Institut f\"ur Kernphysik, Goethe-Universit\"at Frankfurt am Main, Max-von-Laue-Str.1, 60438 Frankfurt, Germany}

\author{\firstname{J.} \surname{Titze}}
\affiliation{Institut f\"ur Kernphysik, Goethe-Universit\"at Frankfurt am Main, Max-von-Laue-Str.1, 60438 Frankfurt, Germany}

\author{\firstname{T.} \surname{Jahnke}}
\affiliation{Institut f\"ur Kernphysik, Goethe-Universit\"at Frankfurt am Main, Max-von-Laue-Str.1, 60438 Frankfurt, Germany}

\author{\firstname{A.} \surname{Czasch}}
\affiliation{Institut f\"ur Kernphysik, Goethe-Universit\"at Frankfurt am Main, Max-von-Laue-Str.1, 60438 Frankfurt, Germany}

\author{\firstname{M.S.} \surname{Sch\"offler}}
\affiliation{Institut f\"ur Kernphysik, Goethe-Universit\"at Frankfurt am Main, Max-von-Laue-Str.1, 60438 Frankfurt, Germany}\affiliation{Lawrence Berkeley National Laboratory, 1 Cyclotron Road, Berkeley, CA-94720, USA}

\author{\firstname{K.} \surname{Kreidi}}
\affiliation{Institut f\"ur Kernphysik, Goethe-Universit\"at Frankfurt am Main, Max-von-Laue-Str.1, 60438 Frankfurt, Germany}

\author{\firstname{O.} \surname{Jagutzki}}
\affiliation{Institut f\"ur Kernphysik, Goethe-Universit\"at Frankfurt am Main, Max-von-Laue-Str.1, 60438 Frankfurt, Germany}

\author{\firstname{H.} \surname{Schmidt-B\"ocking}}
\affiliation{Institut f\"ur Kernphysik, Goethe-Universit\"at Frankfurt am Main, Max-von-Laue-Str.1, 60438 Frankfurt, Germany}

\author{\firstname{R.} \surname{D\"orner }}
\email{doerner@atom.uni-frankfurt.de}
\affiliation{Institut f\"ur Kernphysik, Goethe-Universit\"at Frankfurt am Main,
Max-von-Laue-Str.1, 60438 Frankfurt, Germany}


\date{\today}

\begin{abstract}
Fragmentation of highly charged molecular ions or clusters consisting of more than two atoms can proceed in an onestep synchronous manner where all bonds break simultaneously or sequentially by emitting one ion after the other. We separated these decay channels for the fragmentation of CO$_{2}^{3+}$ ions by measuring the momenta of the ionic fragments. We show that the total energy deposited in the molecular ion is a control parameter which switches between three distinct fragmentation pathways: the \textit{sequential} fragmentation in which the emission of an O$^+$ ion leaves a rotating CO$^{2+}$ ion behind that fragments after a time delay, the Coulomb explosion and an in-between fragmentation - the \textit{asynchronous} dissociation. These mechanisms are directly distinguishable in Dalitz plots and Newton diagrams of the fragment momenta. The CO$_{2}^{3+}$ ions are produced by multiple electron capture in collisions with 3.2 keV/u Ar$^{8+}$ ions.\\

\end{abstract} 
\maketitle\pagebreak

As one or more electrons are removed from a neutral diatomic or polyatomic molecule or cluster the Coulomb repulsion between the ionic cores will eventually lead to fragmentation. The dynamics of this dissociation of molecular and cluster ions, however, is highly complex.  A key question is, what parameters control which of the various decay mechanisms becomes active. In an ideal case, if such parameters are unveiled, they can be used to switch between the breakup channels. The most simple of such mechanisms  is single step Coulomb explosion \cite{Berg1994,Hsieh1997,Singh2006} of the multiply charged molecular ion. Here all bonds break simultaneously and the remaining atomic ions are driven apart purely by their Coulomb repulsion. This scenario is in many cases successfully used to describe fragmentation of molecules in strong laser fields \cite{Cornaggia1994,Gagnon2008,Hasegawa2001}. The other extreme are \textit{sequential} or stepwise processes in which in the first step the molecular ion separates into two fragments. After that - at distances between the primary fragments where these hardly interact anymore - another dissociation takes place. In between these two scenarios a third one is possible - the \textit{asynchronous} breakup: the bonds of the molecular ion break in a single step, but at a time where the geometry of the molecule is asymmetric, as vibration and rotation precede the fragmentation.

In the present study we show for the prototype system of CO$_{2}^{3+}$ ions that these different dissociation pathways can be identified and distinguished experimentally and, furthermore, that the system can be steered on one or the other pathway by varying the total amount of energy deposited into the molecular ion. The decay of the dication of CO$_{2}$ molecules created by electron impact, photoionization or heavy ion impact has been studied theoretically and experimentally by several groups \cite{Bapat2007,Sharma2007,Hochlaf1998,Hogreve1998,King2008,Mrazek2000,Tian1998a}. First studies, including the dissociation of CO$_{2}^{q+}$ with charge states up to q=4, using time-of-flight mass spectrometers were done by Tian \textit{et al.} \cite{Tian1998b} in 1998. Up to present there are only a few more experimental reports on the three-body dissociation of CO$_{2}^{3+}$ ions \cite{King2008}. In last decade, sophisticated experimental techniques to measure not only the time-of-flight of each ionic fragment but also the complete vector momentum gave a better understanding of dissociation processes and molecular geometries, see e.g. \cite{Adoui2001}.

In the present work we have chosen impact of slow highly charged ions (3.2 keV/u Ar$^{8+}$ projectiles) to produce CO$_{2}^{3+}$ ions. The multiple capture of electrons by highly charged ions is a rather gentle and very rapid process which leaves the molecular ion preferentially  in its ground or low lying electronically excited state. Compared to ionization in a femtosecond laser field \cite{Hishikawa1999}, the collision is very fast and does not leave time for geometrical rearrangement. It leads to a vertical transition between the linear neutral ground state and the CO$_{2}^{3+}$ potential energy surfaces. Unlike earlier experiments \cite{Sanderson1999,Adoui2001,Adoui2006,Bapat2007,Sharma2007} we are able to observe the three-body dissociation of CO$_{2}^{3+}$ leading to C$^{+}$+O$^{+}$+O$^{+}$ ions in a kinematically complete way by applying multicoincidence momentum imaging techniques, explained elsewhere \cite{Doerner2000,Ullrich2003,Jahnke2004}. The detection of all fragments with 4$\pi$ solid angle allows us to distinguish the various fragmentation mechanisms of the trication in unprecedented detail and completeness. The ion beam of Ar$^{8+}$ projectiles at 3.2 keV/u has been generated at the Electron Cyclotron Resonance (ECR) Ion Source at Goethe-University, Frankfurt. All three fragment ions are measured in coincidence with the projectile charge state using the COLTRIMS (COLd Target Recoil Ion Momentum Spectroscopy) technique \cite{Doerner2000,Ullrich2003,Jahnke2004}. The ionic fragments produced in the interaction region are guided by an electrical field of 39 V/cm onto a microchannel plate detector with delay-line anode \cite{Jagutzki2002}. By measuring the positions of impact and the time-of-flight of each particle one can determine the three-dimensional initial momentum vector and the mass to charge ratio of each ionic fragment in an offline analysis. Downstream of the reaction zone the projectile charge states are separated by an electrostatic deflector. These projectile ions are detected by a time- and position sensitive microchannel plate detector, as well.\\

\begin{figure}
\centering
\includegraphics[width=8cm]{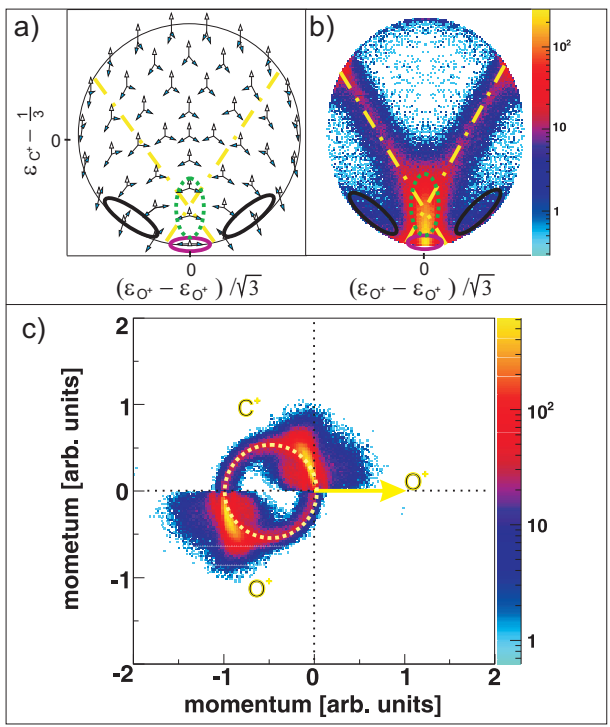}
\caption[map]{(a) Characteristic momentum vector geometries for specific points in the Dalitz plot and (b) Daltiz plot with measured data; (a)~+~(b) events located in the different marked areas correspond to various reaction mechanisms: magenta colored oval - \textit{direct ionization process}, dash-dotted X-shape - \textit{sequential breakup}, black ovals on left and right side - \textit{asynchronous stretching} and green dotted area - \textit{molecular bending}. (c) Newton diagram: momentum vector of one O$^{+}$ ion in the CM frame defines the x-axis, while the relative momentum vectors of the C$^{+}$ ion and the second O$^{+}$ ion are mapped in the upper and lower half, respectively; the dashed circle marks the \textit{sequential} breakup, the symmetric islands in the upper and lower half identify the \textit{direct} and \textit{concerted} breakup mechanism, see text.}
\label{fig:map}
\end{figure}

We now show how we experimentally identify the various breakup mechanisms. A very useful tool for visualization of three body processes is the Dalitz plot \cite{Dalitz1953}. This probability-density plot displays the vector correlation in terms of the reduced energies of the three atoms,

\begin{equation}
\label{eq:daltiz-coord}
        \epsilon_{_{C^{+}}}\,-\,\frac{1}{3}~~~~vs.~~~~\frac{\epsilon_{_{O^{+}}}\,-\,\epsilon_{_{O^{+}}}}{\sqrt{3}}
\end{equation}

where $\epsilon_{_{C^{+},O^{+}}}\,=\,\vec{k}_{_{C^{+},O^{+}}}/(2\,m_{_{C,O}}\,W)$, m is the mass and W is the total energy of the three atoms. A key advantage of the Dalitz plot is, that the phase space density is constant, i.e. all structure in such a plot results from the dynamics of the process, not from the trivial final state phase space density. Figure \ref{fig:map}(b) shows a Dalitz plot of our measured data. Each region of that diagram refers to a certain geometry of the momentum vectors at the instance of breakup as shown in fig. \ref{fig:map}(a). Our data clearly shows that the most likely configuration of the dissociating CO$_{2}^{3+}$ ion is linear (marked by the small violet oval at the bottom). For events in this region the energy of the C$^{+}$ ion is very small and the two O$^{+}$ ions are emitted back-to-back, reflecting the linear ground state geometry of the CO$_{2}$ molecule. This island corresponds to the direct, synchronous process where all bonds break simultaneously.

\begin{figure*}
\centering
\includegraphics[width=1.0\textwidth]{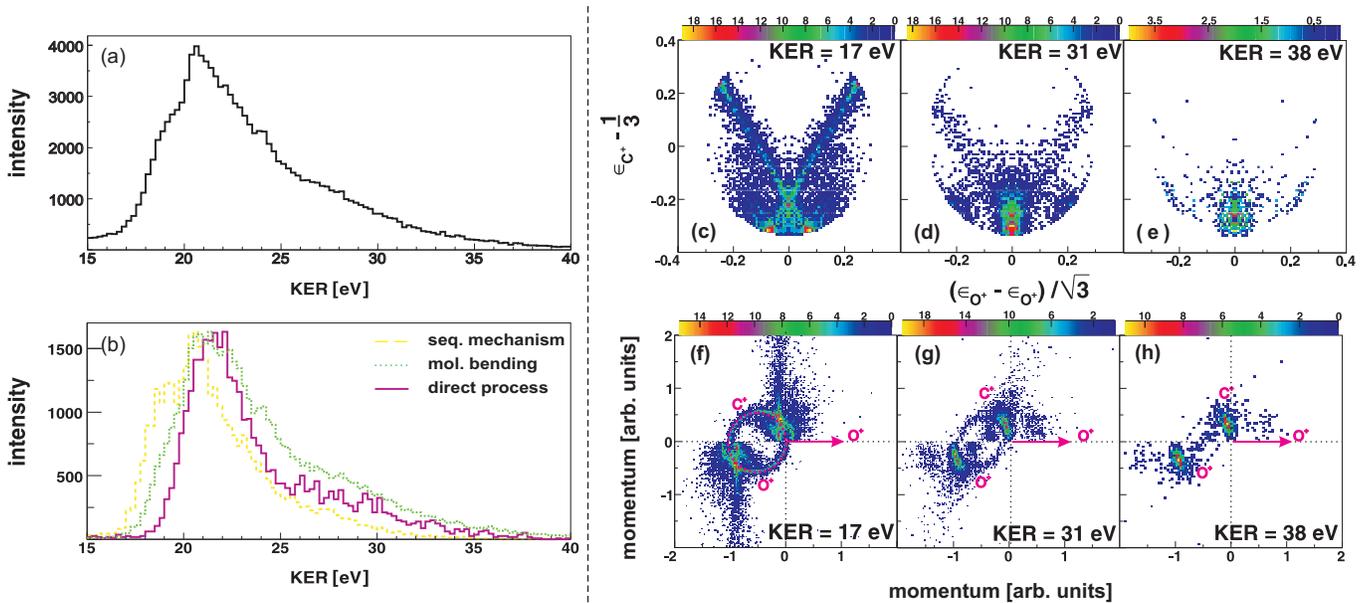}
\caption[newton]{left: (a) Total, experimental KER (black solid line) distribution. (b) KER distributions for the various reaction mechanisms normalized to the maximum, respectively; The different colors represent events selected from different regions in fig. \ref{fig:map}(a) corresponding to the different mechanisms; right: Spectra (c)~-~(e) show Dalitz plots and spectra (f)~-~(h) Newton diagramms for different regions of KER: (c) and (f): 16,5~eV~$<$~KER~$<$~17,5~eV, (d) and (g): 30,5~eV~$<$~KER~$<$~31,5~eV and (f) and (h): 37,5~eV~$<$~KER~$<$~38,5~eV.}
\label{fig:newton}
\end{figure*}

In recent experiments on slow and swift heavy ion collisions with polyatomic molecules like CO$_{2}$ only this simultaneous breakup and some contribution from the asynchronous reaction mechanism have been observed \cite{Sanderson1999, Adoui2006}. Figure \ref{fig:map}(b) clearly proves that in slow ion collisions the asynchronous dissociation mechanism, preceded by molecular bending and asymmetric stretching of the molecular ion, take place, as well; events resulting from molecular bending are located within the green dashed oval, while events allocated to asymmetric stretching can be found inside the black solid-line ovals (left and right at the bottom). Additionally, the Dalitz plot (fig. \ref{fig:map}(b)) shows a fourth, X-shaped structure marked by the yellow dash-dotted lines which contains about 20\% of all events. This structure results from the sequential breakup.
To see this more directly, we display the same data in a Newton diagram in fig. \ref{fig:map}(c). Here the momentum vectors are shown with respect to the center of mass (CM) of the fragments. The direction of the momentum vector of one of the O$^{+}$ ions is represented by an arrow fixed at 1 a.u.. The momentum vectors of the C$^{+}$ ion and the second O$^{+}$ ion are normalized to the length of the first O$^{+}$ ion momentum vector and mapped in the upper and lower half of the plot, respectively. In the Newton diagram the C$^{+}$ and the second O$^{+}$ fragment momenta are located on a circle shifted to the left by about half of the first O$^{+}$ ion momentum. This shifted circle (marked by the dashed line in fig. \ref{fig:map}(c)) is a clear proof of a two step dissociation mechanism. In the first step the CO$_{2}^{3+}$ molecule dissociates into an O$^{+}$ ion and an CO$^{2+}$ ion. Then after a delay, when the force of the departing O$^{+}$ ion onto the CO$^{2+}$ fragment is negligible, the second step occurs: the CO$^{2+}$ ion, which is moving to the left, dissociates into a C$^{+}$ ion and a second O$^{+}$ ion. The intermediate CO$^{2+}$ fragment receives some angular momentum as the O$^{+}$ ion is expelled. From the mean angle of 170$\deg$ for the ground state of CO$_{2}$ and a measured momentum of about 150 a.u. of the primary O$^{+}$ ion the angular momentum transfer to the CO$^{2+}$ ion left behind can be estimated to be about 60~$\hbar$ corresponding to 89 fsec for half a turn. The secondary breakup of this rotating CO$^{2+}$ wavepacket leads to the observed circle in figure \ref{fig:map}(c). Previous studies of CO$^{2+}$ molecules created by K-shell photoionization followed by Auger decay have shown that for a kinetic energy release (KER) below 10.95 eV, the CO$^{2+}$ ion decays within 30-100 fsec, which is sufficient for rotation to occur before fragmentation (see figure 4 in \cite{Weber2001}). One indication for this lifetime is also the clearly observed vibrational structure in this KER regime (see figure  3 in \cite{Weber2001}). The most important intermediate states of the excited CO$^{2+}$ molecule in this regime are the $^1\Pi$, $^3\Sigma^+$ and $2^1\Sigma^+$ \cite{Lundqvist1995,Weber2001,Weber2003}. The potential energy curves of these states have local minima in the 1.9-3.8~a.u. range which decay by coupling to purely repulsive states \cite{Weber2003,Kerkau2001}.

Besides the illustrated sequential dissociation mechanism concerted breakups, like the asynchronous decay, can be identified in the Newton diagram, as well. The geometrical rearrangement of the CO$_{2}^{3+}$ ion after ionization via electron capture results in a momentum gain of the C$^{+}$ fragment. This momentum gain leads to the apparent slight bend angle of the main spots in figure \ref{fig:map}(c) which correspond to the data in the Daltiz plot, associated with the bending and asymmetric stretching mode (area inside the green dashed and black solid ovals in figure \ref{fig:map}(b)). Unlike in this asynchronous reaction mechanism the C$^{+}$ ion gains almost no momentum while breaking up via pure direct processes. Here the O$^{+}$ ions dissociate back-to-back leaving the C$^{+}$ ion almost at rest. This simultaneous breakup leads to islands in the upper and lower half of the Newton diagram, respectively. The slight offset of these main spots is an artefact of the Newton diagram. By definition all carbon ions are displayed in the upper hemisphere and all oxygen ions appear in the lower half. Thus, any spread of a linear configuration looks like an apparent bend. This effect is further enhanced by the fact that unlike the Dalitz plot the Newton diagram does not have a constant phase space: The solid angle and hence the phase space along the horizontal separating the carbon ion from the oxygen ion region is zero.\\

After unambigously identifying the fragmentation pathways we now show, that the amount of energy deposited into the system by the ion impact, decides which pathway is dominant. This energy is converted to kinetic energy of the fragments, which we measure, and to a much smaller extent in possible electronic excitation energy, which eventually is emitted as photons. In our setup we measure the KER with a resolution of about 100 meV, we do not detect emitted photons. Since our multiple electron capture reaction typically does not create very highly excited states of CO$_{2}^{3+}$, mainly, fragments in the ionic ground state contribute in our case. Figure \ref{fig:newton}(a) shows the total KER distribution and figure \ref{fig:newton}(b) shows the measured KER distributions for the different regions in the Dalitz plot indicated in fig. \ref{fig:map}(a) and (b) by the coloured ovals. They correspond to the different breakup mechanisms, as well. The KER distribution for the sequential breakup shows the smallest onset energy of all three mechanisms. Figures \ref{fig:newton}(c)\,-\,(e) show the Dalitz plots and fig. \ref{fig:newton}(f)\,-\,(h) show the Newton diagrams gated on different regions of KER, respectively, i.e. they correspond to different amounts of total energy deposited into the system. Figures \ref{fig:newton}(c) and (f) show the region of energy barely above the threshold for the three body fragmentation. At this threshold, clearly, the fragmentation proceeds predominantly in a two step, sequential fashion. Figures \ref{fig:newton}(d) and (g) show events where an additional energy of 14~eV is brought into the system. Now some flux appears in the region of the black circles in fig. \ref{fig:map}(a) and (b) that corresponds to asymmetric stretching of the molecule before fragmentation. Also some population of the direct breakup channel occurs. For even higher KER (see fig. \ref{fig:newton}(e) and (h)), finally, that direct breakup dominates.

The missing contributions at smaller KERs for simultaneous reaction mechanisms indicate that for energies less than 20 eV above the three body asymptote of the C$^{+}$ +O$^{+}$ +O$^{+}$ final state there are no potential energy surfaces leading to direct breakup within the Franck- Condon region of the CO$_{2}$ molecule. This shows that molecular bending modes can be activated during vertical Franck- Condon transitions not only by fast but also by slow ion impact. We attribute the missing evidence for these modes in previous work \cite{Sanderson1999,Adoui2006} to the much improved resolution and statistics of our present study. Typically, many body fragmentation proceeds via regions in which the potential energy surfaces are very dense and, additionally, many transitions between them are allowed. Predictions based on single dissociation pathways in the multidimensional potential energy landscape, thus, become increasingly impractical. Here our study shows a way out by identifying clear mechanisms directly from the data without the need of knowledge of the potential energy surface. As we have shown, the total energy put into the system is the key parameter which can be used to control the fragmentation.\\


\begin{thebibliography}{27}
\bibitem{Berg1994}
L.~E.~Berg, A. Karawajczyk and C. Stromholm,
\newblock{\em J. Phys. B}, \underline{27}: 2971, 1994


\bibitem{Hsieh1997}
S. Hsieh and J.~H.~D. Eland,
\newblock{\em J. Phys. B}, \underline{30}: 4515, 1997.


\bibitem{Singh2006}
R.~K. Singh  \textit{et al.},
\newblock{\em Phys. Rev. A}, \underline{74}: 022708, 2006


\bibitem{Cornaggia1994}
C. Cornaggia, M. Schmidt and D. Normand,
\newblock{\em J. Phys. B}, \underline{27}: L123, 1994.


\bibitem{Gagnon2008}
J. Gagnon \textit{et al.},
\newblock{\em J. Phys. B}, \underline{41}, 215104, 2008.


\bibitem{Hasegawa2001}
H. Hasegawa, A. Hishikawa and K. Yamanouchi,
\newblock{\em Chem. Phys. Lett.}, \underline{349}, 57, 2001.


\bibitem{Bapat2007}
B. Bapat and V. Sharma,
\newblock{\em J. Phys. B}, \underline{40}: 13, 2007


\bibitem{Sharma2007}
V. Sharma \textit{et al.},
\newblock{\em J. Phys. Chem.}, \underline{111}: 10205, 2007


\bibitem{Hochlaf1998}
M. Hochlaf \textit{et al.},
\newblock{\em J. Phys. B}, \underline{31}: 2163, 1998


\bibitem{Hogreve1998}
H. Hogreve,
\newblock{\em J. Phys. B}, \underline{28}: L263, 1998


\bibitem{King2008}
S.~J. King and S.~D. Price,
\newblock{\em Int. J. Mass. Spectr.}, \underline{272}: 154, 2008


\bibitem{Mrazek2000}
L. Mrazek \textit{et al.},
\newblock{\em J. Phys. Chem. A}, \underline{104}: 7294, 2000


\bibitem{Tian1998a}
C. Tian and C.~R. Vidal,
\newblock{\em J. Chem. Phys.}, \underline{108}: 927, 1998


\bibitem{Tian1998b}
C. Tian and C.~R. Vidal,
\newblock{\em Phys. Rev. A}, \underline{58}: 3783, 1998


\bibitem{Adoui2001}
L. Adoui \textit{et al.},
\newblock{\em Physica Scripta}, \underline{T92}: 89, 2001


\bibitem{Hishikawa1999}
A. Hishikawa, A. Iwamae and K. Yamanouchi,
\newblock{\em Phys. Rev. Lett.}, \underline{83}: 1127, 1999


\bibitem{Sanderson1999}
J.~H. Sanderson \textit{et al.},
\newblock{\em Phys. Rev. A}, \underline{59}: 4817, 1999


\bibitem{Adoui2006}
L. Adoui \textit{et al.},
\newblock{\em Nucl. Instr. and Meth. in Phys. Res. B}, \underline{245}: 94, 2006


\bibitem{Doerner2000}
R. D\"orner\textit {et al.},
\newblock{\em Physics Reports}, \underline{330}: 95, 2000


\bibitem{Ullrich2003}
J. Ullrich \textit{et al.},
\newblock {\em Rep. Prog. Phys.}, \underline{66}: 1463, 2003


\bibitem{Jahnke2004}
T. Jahnke \textit{et al.},
\newblock {\em J. Elec. Spec. and Rel. Phen.}, \underline{141}: 229, 2004


\bibitem{Jagutzki2002}
O. Jagutzki \textit{et al.},
\newblock {\em Nucl.Instr. Meth. A}, \underline{477}: 244, 2002


\bibitem{Dalitz1953}
R.~H. Dalitz,
\newblock{\em Phil. Mag.}, \underline{44}: 1068, 1953


\bibitem{Weber2001}
Th. Weber \textit{et al.},
\newblock{\em J. Phys. B}, \underline{34}:3669, 2001


\bibitem{Lundqvist1995}
M. Lundqvist \textit{et al.},
\newblock{\em Phys. Rev. Lett.}, \underline{75}: 1058, 1995


\bibitem{Weber2003}
Th. Weber \textit{et al.},
\newblock{\em Phys. Rev. Lett.}, \underline{90}: 153003, 2003


\bibitem{Kerkau2001}
T. Kerkau and V. Schmidt,
\newblock{\em J. Phys. B}, \underline{34}: 839, 2001


\end{thebibliography}
\end{document}